\documentstyle{article}
\begin{document} 
\newcommand{\beq}{\begin{equation}}
\newcommand{\eeq}{\end{equation}} 
\newcommand{\bea}{\begin{eqnarray}}
\newcommand{\eea}{\end{eqnarray}} 
\newcommand{\nn}{\nonumber}
\newcommand{\bal}{\begin{array}{ll}} 
\newcommand{\eal}{\end{array}}
\newcommand{\la}[1]{\lambda^{#1}}
\def\1{{\rm 1 \kern -.10cm I \kern .14cm}} \def\R{{\rm R \kern -.28cm I
\kern .19cm}}

\begin{titlepage} 
\begin{flushright}  LPTHE-ORSAY 95/54 \\UFIFT-HEP-96-1 \\ hep-ph/9601243  
\\January 1996
\end{flushright} 
\vskip .8cm 
\centerline{\LARGE{\bf {Yukawa textures with an anomalous}}} 
\vskip .5cm
\centerline{\LARGE{\bf {horizontal abelian symmetry}}}   
\vskip 1.5cm \centerline{\bf {Pierre Bin\'etruy and St\'ephane Lavignac
\footnote{Supported in part by the Human Capital and Mobility
Programme, contract CHRX-CT93-0132.}}}  
\vskip .5cm   \centerline{\em
Laboratoire de Physique Th\'eorique et Hautes Energies
\footnote{Laboratoire associ\'e au CNRS-URA-D0063.}} \centerline{\em
 Universit\'e Paris-Sud, B\^at.
211,} \centerline{\em F-91405 Orsay Cedex, France }
\vskip 1cm \centerline{\bf {Pierre Ramond}\footnote{Supported in part
by the United States Department of Energy under grant
DE-FG05-86-ER40272.}} \vskip .5cm \centerline{\em Institute for
Fundamental Theory,}  \centerline{\em Department of Physics,
University of Florida} \centerline{\em Gainesville FL 32611, USA} 
\vskip 1cm \centerline{\bf {Abstract}}

\indent

The observed hierarchy of quark and lepton masses and mixings may be obtained
by adding an abelian family symmetry to the Minimal Supersymmetric Model and
coupling quarks and leptons to an electroweak singlet scalar field. In a
large class of such models, this symmetry suffers from anomalies which must
be compensated by the Green-Schwarz mechanism; this in turn fixes the
electroweak mixing angle to be $\sin^2 \theta_W = 3/8$ at the string scale,
without any assumed GUT structure. The analysis is extended to two distinct
generalisations of the Standard Model: neutrino masses and mixings and
R-parity violating interactions. 

\end{titlepage} 

\section{Introduction.} \label{sec:nom1} 

The problem of quark and lepton mass hierarchies and mixing is not addressed
by the Standard Model and has been a thorn in particle theorists side. Recent
developments, both experimental and theoretical, might shed new light on this
long standing issue. On the  experimental side, it is the discovery of the top
quark \cite{top} in the mass range of the electroweak scale: in more
technical terms, the top Yukawa coupling is found to be of the order of the
gauge couplings. On the theoretical side, the emergence of string theories as
a universal theory encompassing all known fundamental interactions including
gravity provides a unique framework which allows to relate features of the
effective low energy theory which seemed heretofore uncorrelated. Of special
interest for the problem that we are discussing are: the presence of
non-renormalisable interactions (which can in principle be computed within a
given string model); an often large number of horizontal gauge symmetries,
especially abelian, which are spontaneously broken at scales which may vary
between the electroweak scale and the Planck scale; a large number of Standard
Model singlet scalar fields whose couplings to ordinary matter are fixed by the
latter symmetries.

All these properties have induced several groups to reconsider the original
idea of Froggatt and Niesen \cite{FN} which uses nonrenormalisable couplings of
quarks and leptons to electroweak gauge single fields and an
horizontal symmetry to constrain these couplings in order to generate mass
hierarchies.The first results are promising and lead to new theoretical
developments and new ways to test experimentally these ideas. 

We address some aspects of this program in this paper. In section 2, we
recall the basic concepts and stress the relevance of some parameters, such
as the supersymmetric $\mu$-term. We then proceed to discuss the connection
between the phenomenological constraint coming from the quark and lepton mass
spectrum and the more fundamental issue of the anomaly structure of the
horizontal family. We show that, for a large class of models, phenomenology
requires our abelian symmetry to be anomalous, this anomaly being cancelled
by a Green-Schwarz mechanism \cite{GS}. This property obviously points
towards string theories. Section 3 is devoted to the the study of neutrino
masses and mixings when one adds to the particle content of the Standard
Model right-handed neutrinos. It is shown that the abelian horizontal
symmetry provides constraints on the neutrino mass spectrum as well as on
the mixing angle. In section 4, we consider another extension of the
Standard Model: the spectrum remains minimal but we allow for couplings
which violate R-parity. Again, the horizontal symmetry constrains these new
couplings. Finally, section 5 gives our conclusions.


\section {Strategies with chiral scalars.}\label{sec:nom2}

The basic idea, which dates back to the early papers of Froggatt and
Nielsen \cite{FN} is to use an abelian horizontal symmetry $U(1)_X$ in
order to forbid most Yukawa couplings: in practice all  but
the top quark coupling or all but the third family couplings. The
hierarchies of fermion masses and mixings are then generated through
higher dimensional operators involving one or several electroweak singlet
scalar fields. These fields acquire a vacuum expectation value which
breaks the horizontal symmetry at some large scale and gives rise to the
ordinary Yukawa couplings. More specifically, if $\theta$ is one such
field of $X$-charge $-1$, $X$-charge conservation allows for example the
non-renormalisable term in the superpotential: 
\beq
\lambda^U_{ij} Q_i \bar u_j H_u \left( {\theta \over M} \right)^{n_{ij}}
\label{eq:generic} 
\eeq
where $Q_i$ is the quark isodoublet of the $i$th generation, $\bar u_j$
is the u quark-type isosinglet of the $j$-th generation, $H_u$ is one
of the two Higgs doublets of the supersymmetric standard model. The
coupling $\lambda^U_{ij}$ is expected to be of order one and the mass
$M$ is a large mass scale, the order of which we will discuss later. The
positive rational number $n_{ij}$ is nothing but the sum of the
$X$-charges of the standard model fields involved, namely $Q_i$, $\bar
u_j$ and $H_u$:
\beq
n_{ij} = q_i + u_j + h_u. \label{eq:inxs}
\eeq

Once $\theta$ gets a vacuum expectation value, one obtains an effective
Yukawa coupling:
\beq
Y^U_{ij} = \lambda^U_{ij} \left( {<\theta> \over M} \right)^{n_{ij}} . 
\label{YU} 
\eeq
If $<\theta> / M$ is a small number, and if the array of $X$-charges is
sufficiently diversified, one may implement in the theory various
hierarchies of masses and mixings. Our goal is to select a class of
models where such a strategy proves to be efficient as well as it
leads to specific predictions. In this Section, we will review the
possibilities that are open to us in order to decide which lead to the
most interesting and fruitful properties. 

The electroweak singlet fields $\theta$ may appear in vectorlike pairs
or as chiral individuals. In the latter case, the low energy Yukawa
matrix will contain zeroes whenever the excess charge $n_{ij}$ turns
out to be negative, since, in this case, the holomorphy of the superpotential 
prevents \cite{NLS} a coupling of the type (\ref{eq:generic}): we will thus 
refer to them as {\em supersymmetric zeroes}. Such a property  may or may
not be a welcome feature, since it may yield too many zeroes in the
mass matrix. One may thus prefer to introduce a vectorlike pair
$(\theta, \bar \theta)$ of electroweak singlets of respective
$X$-charge $(-1)$ and $(+1)$. If they correspond to D-flat directions,
then naturally $<\theta> = <\bar \theta>$ and the low energy Yukawa
couplings will be of order  $ (<\theta> / M)^{|n_{ij}|}$, irrespective
of the sign of the excess charge $n_{ij}$ \cite{IR}. The problem with
this approach is that a supersymmetric mass term $\tilde M \theta \bar
\theta$ is perfectly allowed by the symmetries (unless one assumes an unwelcome
fine tuning, $\tilde M$ is a large mass scale) and spoils the D-flat
direction, leading to a large hierarchy between the vacuum expectation
values.

On the other hand, we have shown \cite{BR} that, in a large class of
models with a chiral $\theta$ field, there exists an interesting
connection between the fermion mass spectrum and the value of the
Weinberg angle. More precisely, the fermion mass spectrum puts such
constraints on the $X$-charges that the mixed anomalies of the $U(1)_X$
symmetry are necessarily nonzero and must be cancelled using the
Green-Schwarz mechanism \cite{GS}. As generically stressed by
Ib\'a\~{n}ez \cite{Ib}, this in turn fixes also the weak mixing angle
which we find equal to its standard value of $\sin^2 \theta_W = 3/8$ at
the superheavy scale. We will return to this question below but this
attractive feature leads us to concentrate in the rest of this paper on
the class of models with only chiral electroweak singlet scalars ({\em
i.e.} no vector-like pair).

\subsection{Filling the supersymmetric zeroes through wave function 
renormalisation}

It has been stressed before \cite{NLS,DPS} that, in this class of
models, non-renormalizable contributions to the fermion kinetic terms
may lead to filling the zeroes imposed by supersymmetry
(corresponding to $n_{ij} <0$). Let us take this opportunity to discuss
our general strategy. We are considering  the effective theory obtained
from a more fundamental theory  of typical scale $M$, well below this
scale $M$. The fields of the effective theory are, by assumption, those
of the Minimal Supersymmetric Model plus electroweak singlet chiral
scalars, generically denoted as $\theta$. We are  writing the most
general couplings  including non-renormalizable terms proportional to
negative powers of $M$, compatible with the symmetries of the effective
theory, namely $SU(3) \times SU(2) \times U(1)_Y \times U(1)_X$.  This
yields terms of the type (\ref{eq:generic}), and similar couplings for
the charge $(-1/3)$ quarks and charge $(-1)$ leptons.  It may also give
rise to $R$-parity breaking interactions. We will study this possibility
in Section  \ref{sec:nom4}.  Our concern here is that it also gives
kinetic terms for the  fermions with a $\theta$ dependent normalisation.
The low energy fermion fields are therefore obtained through a $\theta$
dependent redefinition, which may modify the $\theta$ dependence of
the Yukawa couplings.

For concreteness, let us consider the Yukawa couplings arising from
(\ref{eq:generic}). The normalized kinetic terms originate from a
diagonal quadratic K\"ahler potential of the form
\beq
K_0(Q_i,\bar u_j,\cdots) = Q_i^+ Q_i + \bar u_i^+ \bar u_i + \cdots  
\eeq
In our case, the K\"ahler potential as well receives non-renormalisable 
contributions; it reads, to lowest order in powers of $1/M$:
\bea
K(Q_i,\bar u_j,\cdots,\theta) &=& Q_i^+ Q_j \left[ H(q_i - q_j)
\left( {\theta^+ \over M} \right)^{q_i - q_j} + H(q_j - q_i) \left(
{\theta \over M} \right)^{q_j - q_i} \right] \nonumber \\
&+& \bar u_i^+ \bar u_j \left[ H(u_i - u_j) \left(
{\theta^+ \over M} \right)^{u_i - u_j} + H(u_j - u_i) \left(
{\theta \over M} \right)^{u_j - u_i} \right] \nonumber \\
&+& \cdots   \label{eq:kinetic}
\eea
where $H(x)$ is the Heaviside function ($H(x) = x$ if $x \ge 0$, $H(x) = 0$
otherwise). To bring the kinetic terms to their canonical form, we have
to redefine the matter fields $\Phi_i$ ($\Phi = Q, \bar u, \bar d, L,
\bar e$):
\beq
\Phi_i \rightarrow V^{\Phi}_{ij} \Phi_j
\eeq
where the order of magnitude of the matrix elements of $V^\Phi$ depends
on the relative charges $\phi_i$ of the $\Phi_i$ fields:
\beq
V^{\Phi}_{ij} \sim \left( {<\theta> \over M}
\right)^{|\phi_i-\phi_j|}.
\eeq
It is useful to note that the structure ot the matrix $V_{\Phi}$ is
simply that of the identity matrix corrected by positive powers of 
$<\theta> / M$.

The Yukawa couplings in the canonical basis 
\beq
{\hat Y}^U = V^{QT} Y^U V^{\bar u}
\eeq
are now a sum of terms
\beq
{\hat Y}^U_{ij}= \sum_{kl} Y_{ij,kl} \label{eq:Sum}
\eeq
where
\beq
Y_{ij,kl} \sim H(q_k + u_l + h_u) \left( { <\theta> \over M }
\right)^{|q_i-q_k|+|u_l-u_j|+q_k+u_l+h_u}. \label{eq:Yijkl}
\eeq
One immediately infers that ${\hat Y}^U_{ij}$ is at most of the order of
magnitude that would be obtained with a vectorlike pair of $\theta$ fields: 
$(<\theta> / M)^{|n_{ij}|}$. This means that, as far as hierarchies are 
concerned, one does not gain much by going to a vectorlike pair scenario, 
the weaknesses of which we stressed earlier.

In the case where $n_{ij} \ge 0$, one deduces from (\ref{eq:Yijkl}) that
\beq
{\hat Y}^U_{ij} \sim \left( <\theta> \over M \right)^{n_{ij}}.
\eeq
In other words, non-zero entries to the Yukawa matrix are left untouched 
by the process of normalizing the kinetic terms.

On the other hand, in the case where $n_{ij} < 0$, one can easily show from 
(\ref{eq:Yijkl}) that
\beq
Y_{ij,kl} = H(n_{kl}) \left( <\theta> \over M \right)^{
|n_{ij}| + 2max(n_{kl}, n_{kj}, n_{il})}, \label{eq:Yijklneg}
\eeq
which shows that ${Y^U_{ij}}'$ is of order $(<\theta>/M)^{|n_{ij}|}$ or 
smaller.

As an example, we can apply the above results to the case where the (3,3)
entry to the Yukawa matrix is allowed by the $U(1)_X$ symmetry, {\em i.e.}
$n_{33} = 0$. Then applying (\ref{eq:Yijklneg}) with all indices equal to $3$
except for either $i$ or $j$, one finds
\beq
{\hat Y}^U_{i3} \sim \left( <\theta> \over M \right)^{|n_{i3}|}, \;\;
{\hat Y}^U_{3j} \sim \left( <\theta> \over M \right)^{|n_{3j}|}.
\eeq
Similarly, for $i$ and $j$ different from 3, if both $n_{i3}$ and $n_{3j}$
are negative
\beq
{\hat Y}^U_{ij} \sim \left( <\theta> \over M \right)^{|n_{ij}|}.
\eeq
Since $n_{ij} = n_{i3} + n_{3j}$, the corresponding zero in the original  
Yukawa matrix results in this case from the simultaneous presence of zeroes 
in the third line ($n_{3j} < 0$) and third column ($n_{i3} < 0$). If, on the 
other hand, only one is negative, say $n_{i3} < 0$, $n_{3j} \ge 0$, then one
shows that
\beq
{\hat Y}^U_{ij} \sim \left( <\theta> \over M \right)^{|n_{ij}| + 2 min 
(n_{i'j}, n_{ij'}, n_{3j})}.
\eeq
where $i' \not= i \not= 3$ and $j' \not= j \not= 3$ and one used the fact that 
$det \ Y_U \not= 0$.

\subsection{Horizontal abelian charges and the quark and lepton masses}

As introduced in Ref. \cite{BR}, the most general assignment for an Abelian 
horizontal charge to the particles of the Supersymmetric Standard Model reads
\beq
X = X_0 + X_3 + \sqrt{3} X_8, \label{eq:X}
\eeq
where $X_0$ is the family-independent part, $X_3$ is along $\lambda_3$,
and $X_8$ is along $\lambda_8$, the two diagonal Gell-Mann matrices of
the $SU(3)$ family space in each charge sector. In a basis where the
entries correspond to the components in the family space of the fields
$Q$, $\overline u$, $\overline d$, $L$, and $\overline 
e$, we can write the different components in the form
\beq
X_i^{}=(a^{}_i,b^{}_i,c_i^{},d_i^{},e_i^{})\ ,\label{eq:Xfamily}
\eeq
for $i=0,3,8$.
The Higgs doublets $H_{u,d}$  have  X-charges $h_u$ and $h_d$ respectively.
These could be assumed to be equal since, 
using $U(1)_Y$, we have the freedom to redefine the horizontal symmetry in 
order to make these two X-charges equal.  We will return to this later. In any 
case, most of the following discussions depend only on the sum of these 
charges and we thus define
\beq
2h \equiv h_u + h_d. \label{eq:h}
\eeq
Then the excess X-charges $n_{ij}$ defined in (\ref{eq:inxs}) read for the 
charge $2/3$ quarks:
\bea
 & & \left[ {U_0 \over 3} - 2 (a_8+b_8)\right] \left( \begin{array}{lll}
	1 & 1 & 1 \\ 1 & 1 & 1 \\1 & 1 & 1  \end{array}  \right) \nonumber \\
 &+& \left( \begin{array}{lll}
	3(a_8+b_8)+a_3+b_3 & 3(a_8+b_8)+a_3-b_3 & 3a_8+a_3 \\
	3(a_8+b_8)-a_3+b_3 & 3(a_8+b_8)-a_3-b_3 & 3a_8-a_3 \\
	3b_8+b_3 & 3b_8-b_3 & 0   \end{array}  \right)  
\label{eq:xsmatrix}
\eea
and similarly for the charge $-1/3$ quarks with the replacement $(a,b) 
\rightarrow (a,c)$ and for the charge $-1$ leptons with $(a,b) \rightarrow
(d,e)$. In (\ref{eq:xsmatrix}) and the corresponding matrices for the charge 
$-1/3$ and $-1$ sectors, we define the family-independent overall charges:
\bea
U_0 &=& 3(a_0 + b_0 + h_u) \nonumber \\
D_0 &=& 3(a_0 + c_0 + h_d) \label{eq:UDE}  \\
E_0 &=& 3(d_0 + e_0 + h_d). \nonumber
\eea
Some of the excess charges in (\ref{eq:xsmatrix}) might be negative leading 
to supersymmetric zeroes in the Yukawa matrix, to be filled in the way 
described in the previous subsection. But a very generic result, independent to
a large extent of this filling procedure, applies to the determinant of the
Yukawa coupling matrices:
\bea
det\  {\hat Y}^{U} \sim & \left( <\theta> / M_U \right)^{U_0} 
\nonumber \\
det\  {\hat Y}^{D} \sim & \left( <\theta> / M_D \right)^{D_0} 
\label{eq:det} \\
det \  {\hat Y}^{E} \sim & \left( <\theta> / M_E \right)^{E_0}. 
\nonumber 
\eea
The only assumption is that there are not enough supersymmetric zeroes to
make these determinants vanish (hence the $u$ quark
mass is nonzero \cite{BNS}). In these equations, we allowed for different 
scales $M$ in the three different sectors. We will come back to this in a 
later subsection.

The experimental values of the quark and lepton masses,
extrapolated near the Planck scale, satisfy  the order of magnitude
estimates \cite{RRR}
\bea
& & {m_u\over m_t}={\cal O}(\lambda^8)\ ,\;\;\;  {m_c\over m_t}={\cal
O}(\lambda^4)\ , \label{eq:top}\\
& &  {m_d\over m_b}={\cal
O}(\lambda^4)\ , \;\;\;  {m_s\over m_b}={\cal
O}(\lambda^2)\ ,\label{eq:bottom}\\
& &  {m_e\over m_\tau}={\cal O}(\lambda^4)\ , \;\;\;  {m_\mu\over m_\tau}={\cal
O}(\lambda^2)\ ,\label{eq:lept}
\eea
where, following Wolfenstein's parametrization \cite{wolf}, we use the
Cabibbo angle $\lambda$, as expansion parameter. Thus, the mass hierarchy 
appears to be geometrical in each sector. The equality
\beq
m_b=m_\tau\ ,
\eeq
known to be valid in the ultraviolet \cite{btau}, then yields the estimate
\beq
{m_dm_sm_b\over m_e m_\mu m_\tau}={\cal O}(1) \ .\label{eq:goodrat}
\eeq
Of course, all these estimates should be taken with some precaution since
$\lambda$ is not such a small parameter ( thus $2 \lambda^n \sim \lambda^{n-1} 
/2$ ) and the exponents in (\ref{eq:top}-\ref{eq:lept}) should be considered 
as  valid up to a unit. In particular, the ratio $m_e / m_\tau$ is somewhat
closer to  $\lambda^5$ \cite{nir}, which, all other mass ratios being kept
unchanged, gives a ratio (\ref{eq:goodrat}) of order $\lambda$ . We nevertheless find the geometrical hierarchy an 
attractive mass pattern. Comparison of (\ref{eq:goodrat}) with (\ref{eq:det}) 
yields in this case the simple phenomenological constraint:
\beq
D_0 = E_0 \label{eq:geomcons}
\eeq
which, from now on, we will refer to as the {\em geometrical hierarchy 
constraint}. 

Another low energy mass scale which will play an important role in the 
discussion that follows is the so-called $\mu$-term. The origin of such a low 
energy scale in any theory whose fundamental scale is of the order of the 
Planck scale poses problem. The following  solutions have been proposed:

(i) introduce a field $N$  singlet under the Standard Model gauge symmetries 
which  has a trilinear couplings to the Higgs doublets \cite{NSW}:
$\delta W = \lambda N H_u H_d$.

(ii) introduce additional terms in the K\"ahler potential which are quadratic 
in the Higgs fields \cite{GM,KL}:
\beq
\delta K = G(M, M^+) H_u H_d + \; {\rm h.c.}
\eeq
where G is some function of gauge singlet scalars $M$ and their complex 
conjugates $M^+$. If the function $G$ turns out to be some function
analytic in the scalars $M$, then, through a K\"ahler transformation,
this can be rephrased as follows:

(iii) add a nonrenormalisable contribution to the superpotential quadratic in 
the Higgs fields \cite{nilles,CM}:
\beq
\delta W = F(M) H_u H_d.
\eeq

In the context of string models, it is quite plausible that the singlet fields 
involved  are moduli fields which are neutral under the horizontal symmetry 
that we consider. In this case, for any of these scenario to work, we need to 
impose that $h=0$. We will thus refer to it in the sequence as the $h=0$
{\em option}. This was the solution that we adopted in Ref.\cite{BR}.

On the other hand, as emphasized  by Nir \cite{nir} (see also Ref.
\cite{JS2}), the singlet field $\theta$ that we use might provide itself the
solution to the $\mu$-problem \cite{NLS}, following the same scenarios. In
cases (i) and (iii), the following interaction would be allowed by the
horizontal symmetry: \beq
\delta W = M H_u H_d \left( {\theta \over M}  \right)^{2h}
\label{eq:Wmu}
\eeq
where the holomorphy of the superpotential imposes  that $h>0$. 
The $\mu$ term thus obtained is of order
$M(<\theta>/M)^{2h}$ and since, as we will see in subsection 2.5, $M$ is a 
scale close to the Planck scale, one needs a large positive value for $h$.    
  
In case (ii), the K\"ahler potential includes a term\footnote{A similar term 
involving the field $\theta$ itself can be cast into the preceding form 
(\ref{eq:Wmu}), through a K\"ahler transformation; and terms involving both 
$\theta$ and $\theta^+$ are of higher order in $1/M$.}
\beq
\delta K = H_u H_d \left( {\theta^+ \over M}  \right)^{-2h},
\label{eq:mukahler} 
\eeq
which obviously requires $h <0$. The $\mu$ term is then of order $m_{3/2}
(<\theta>/M)^{-2h}$ and thus such an option  works for values of $h$ 
moderately negative.

\subsection{Anomalies}

In Ref.\cite{BR}, we stressed the  important connection between
the anomaly issue and the phenomenological constraints coming from the fermion 
masses. We will repeat the analysis here in the more general framework that we 
have adopted \cite{DPS,nir}.

 The three chiral families
contribute to the mixed gauge anomalies as follows
\bea
C_3&=&3(2a_0^{}+b_0^{}+c_0^{})\ ,\label{eq:anom3}\\
C_2&=&3(3a_0^{}+d_0^{})+ 2h\ ,\label{eq:anom2}\\
C_1&=&a_0^{}+8b_0^{}+2c_0^{}+3d_0^{}+6e_0^{}+2h\ .\label{eq:anom1}
\eea
The subscript denotes the gauge group of the Standard Model, {\it i.e.}
$1\sim U(1)$, $2\sim SU(2)$, and $3\sim SU(3)$. The important feature of 
these three anomaly coefficients is that they depend only on the
family independent charges $X_0$ and thus can be directly related to the 
determinant of the Yukawa matrices through (\ref{eq:UDE},\ref{eq:det}).
The relation depends on the charge $h$ whose connection with the $\mu$
parameter we have stressed in the previous subsection.

The X-charge also has a mixed gravitational anomaly, which is simply, up
to a normalisation,
the trace of the X-charge,
\beq
C_g=3(6a_0^{}+3b_0^{}+3c_0^{}+2d_0^{}+e_0^{})+4h+C_g^\prime\ ,\label{eq:anomg}
\eeq
where $C_g^\prime$ is the contribution from the massless particles that do not
appear in the minimal $N=1$ model.
One must also account for the mixed $YXX$ anomaly, given by
\beq
C_{YXX}=6(a_0^2-2b_0^2+c_0^2-d_0^2+e_0^2)+ 2 (h_u^2 - h_d^2) +4A_T\ ,
\label{eq:anomixed}
\eeq
with the texture-dependent part given by
\beq
A_T=(3a_8^2+a_3^2)-2(3b_8^2+b_3^2)+
(3c_8^2+c_3^2)-(3d_8^2+d_3^2)+(3e_8^2+e_3^2) \ .\label{eq:AT}
\eeq
The last anomaly coefficient is that of the X-charge itself, $C_X$,
the sum of the cubes of the X-charge.

As just emphasized, it is of interest  for our purposes 
that $C_1$, $C_2$, $C_3$ and $C_g - {C_g}'$ only depend on the 
family-independent charges and can thus be related to the determinants of the 
mass matrices through (\ref{eq:det}) \cite{BR}. Indeed, one can easily show 
that the only two independent combinations of these anomaly coefficients 
which can be expressed in terms of $U_0$, $D_0$, $E_0$ and $h$ are
\bea
C_3 &=& (U_0 + D_0) - 6h, \nonumber \\
C_1 + C_2 &=& {8 \over 3} (U_0 + D_0) + 2 (E_0 - D_0) - 12h, \label{eq:combi}
\eea
which involve only $(U_0 + D_0)$ and $(E_0 - D_0)$.

Interesting combinations are $C_1 + C_2 - 8C_3/3$ which depends only on $h$ and
$E_0 -D_0$ and plays a role in the models with a geometrical hierarchy 
\cite{BR}; and $C_1 + C_2 - 2 C_3$ which does not depend on $h$ \cite{DPS,nir}.

It is interesting to express in turn the family independent charges in terms 
of the anomaly coefficients and the Higgs charges:
\bea
a_0 = {\bf +{1 \over 3}} ({D_0 \over 3} - h_d) & {\bf + {1 \over 3}} 
{\cal C}_D & \nonumber \\
b_0 = {\bf -{4 \over 3}} ({D_0 \over 3} - h_d) & {\bf - {1 \over 3}} 
{\cal C}_D  & \; \; \; \; \; \; \; \; \; \; + {1 \over 3} C_3 \nonumber \\
c_0 = {\bf +{2 \over 3}} ({D_0 \over 3} - h_d) & {\bf - {1 \over 3}} 
{\cal C}_D & \nonumber \\
d_0 = {\bf -1 } ({D_0 \over 3} - h_d) & {\bf -1 } 
{\cal C}_D &+{{\bf 1} \over 3} C_2 - {2 \over 3} h  \nonumber \\
e_0 = {\bf +2} ({D_0 \over 3} - h_d) & {\bf +1 } 
{\cal C}_D & - {{\bf 1} \over 3} C_2 + {1 \over 6} (C_1 + C_2 
- {8 \over 3} C_3),
 \label{eq:cd}
\eea
where ${\cal C}_D = -(C_g - {C_g}') /3 + C_1/6 + C_2/2 + 5C_3/9$ 
and we have arranged the right-hand side of these equations such that 
contributions proportional respectively to the $Y$, $B-L$ 
and $L$ charges of the corresponding fields appear in columns.

This shows that one can set $a_0 = -c_0$ by using the $U(1)_Y$ symmetry
to redefine the X charges. In this case, the first column is suppressed and all
charges are expressed in terms of the anomaly coefficients and of the two
Higgs charges (this does not mean of course that $D_0$ can be made to vanish; 
instead we have $D_0 = 3 h_d$).

If the theory also has a $U(1)_{B-L}$ symmetry, one
can further set $a_0 = 0$. Moreover, since the gravitational anomaly 
$C_g - {C_g}'$ is exactly along the $B-L$ charge, one can altogether cancel 
it if one includes a right-handed neutrino to make the  $U(1)_{B-L}$ symmetry 
non-anomalous ({\em i.e. } traceless). 

The parametrisation (\ref{eq:cd}) allows to treat easily the case with no 
mixed gauge anomalies: $C_1 = C_2 = C_3 = 0$. Indeed, one immediately reads off
the charges (with the Y component in the first column subtracted) and deduces 
that $U_0 = 3h_u$, $D_0 = 3 h_d$ and $E_0 = 2h_d - h_u$. Assuming a geometric
hierarchy (\ref{eq:geomcons}) yields $-U_0=D_0=E_0$ ($h=0$) which is easily 
seen not to hold.

We thus turn to the models where the anomaly coefficients are non-zero.
In this case, the anomalies must be cancelled by the Green-Schwarz mechanism
\cite{GS}.
String theories contain an antisymmetric tensor  field which, in 4
dimensions, couples in a universal way to the divergence of the anomalous
currents. One can therefore use the Green-Schwarz mechanism to cancel the
anomalies. Due to the universality of the couplings of this axion-like field,
this is only possible if the mixed anomaly coefficients appear in commensurate
ratios:
\beq
{C_i \over k_i} = {C_X \over k_X} = {C_g \over 12}  \label{eq:GS}
\eeq
where the $k$'s are the Kac-Moody levels at which the corresponding group
structures appear. They are integers in the case of non-abelian
groups and all string models constructed so far have $k_2 = k_3$, which implies
\beq
C_2 = C_3. \label{eq:23}
\eeq
These Kac-Moody levels appear themselves in the gauge coupling unification
condition which is valid at the string scale, without any assumed GUT
structure. This condition reads:
\beq
k_i g_i^2  = k_X g_X^2  = g_{string}^2. \label{eq:unif}
\eeq

As mentioned earlier, one can relate the ratio of d-type quark masses to
charged lepton masses with a combination of anomaly coefficients which can be
turned, using (\ref{eq:GS}), into a combination of Kac-Moody levels, and, using
(\ref{eq:unif}), into a combination of gauge couplings. 

More precisely, using (\ref{eq:UDE},\ref{eq:det}), one obtains, assuming $M_D
= M_E$,
\beq 
{m_d m_s m_b \over m_e m_\mu m_\tau} = {det \hat Y^D \over det \hat Y^E}
=\left( {<\theta> \over M_D }\right)^{3(a_0+c_0-d_0-e_0)}.
\eeq
Hence, through  (\ref{eq:cd}),
\beq 
{m_d m_s m_b \over m_e m_\mu m_\tau}
=\left( {<\theta> \over M_D} \right)^{2h - (C_1 + C_2 - 8/3 C_3)/2}.
\label{eq:masterform}
\eeq

In the $h=0$ option, the geometrical hierarchy discussed above which gives a
mass ratio of order one yields the following relation among anomaly
coefficients \cite{BR}:
\beq
C_1 + C_2 = {8 \over 3} C_3,
\eeq
or, using (\ref{eq:GS}-\ref{eq:unif}),
\beq
{C_1 \over C_2} = {g_2^2 \over g_1^2}= 5/3.
\eeq
This fixes the value of the electroweak angle to its standard GUT value,
without any underlying GUT structure:
\beq
\sin^2 \theta_W = {3 \over 8}.
\eeq

Alternatively, one can start from (\ref{eq:masterform}) and impose the
standard value for the electroweak angle. This is only possible for a
vanishing $h$ in which case one recovers the geometrical hierarchy, or a
moderately negative $h$ (in fact  $h=-1/2$) when one departs slighly from
a geometrical hierarchy ($m_e / m_\tau \sim \lambda^5$) \cite{nir}. As
discussed above, in the latter case, one may use the $\theta$ field to 
account for the $\mu$  term; using (\ref{eq:mukahler}) and
(\ref{eq:masterform}), one obtains
\beq
{m_e m_\mu m_\tau \over m_d m_s m_b}= {\mu \over m_{3/2}}.
\eeq 
The former case necessarily involves  another gauge singlet field
in order to generate a $\mu$ term. 

\subsection{Eigenvalues and mixing angles}

In Ref. \cite{BR}, we presented a result on the hierarchy of mass matrix eigenvalues in models with a vectorlike pair ($\theta$, $\bar \theta$) of singlet scalars. This result can be generalized to the class of models that we are considering in this paper, namely models with a chiral singlet scalar $\theta$. After filling the supersymmetric zeroes, the orders of magnitude of the Yukawa couplings are:
\beq
{\hat Y}_{ij} \sim \left( {<\theta> \over M} \right) ^{\rho_{ij}}
\eeq
where $\rho_{ij}$ is the power of the dominant term in the sum (\ref{eq:Sum}). This hierarchical structure results in a strong hierarchy between the eigenvalues of $Y$. Provided that $\rho_{33} \leq \rho_{ij}$, this hierarchy can be expressed in terms of the two following quantities:
\bea
 p  &  =  &  \min \: (\rho_{11}, \rho_{12}, \rho_{21}, \rho_{22})  \\
 q  &  =  &  \min \: (\rho_{11} + \rho_{22}, \rho_{12} + \rho_{21})
\eea
Normalized to the largest eigenvalue, whose order of magnitude is given by 
${\hat Y}_{33}$, the mass eigenvalues are:
\beq
 \begin{array}{llllll}
 1  &  {\cal O} (<\theta>/M)^{q \over 2}  &  {\cal O} (<\theta>/M)^{q \over 2}  &  &  \mbox{if}  &  p \geq \frac{q}{2}  \\
 \\
 1  &  {\cal O} (<\theta>/M)^{p}  &  {\cal O} (<\theta>/M)^{q-p}  &  &  
\mbox{if}  &  p < \frac{q}{2}
 \end{array}
\label{eq:MassHierarchy}
\eeq
the only case of phenomenological interest being $p < q/2$.

In the simple case studied by Froggatt and Nielsen \cite{FN} where (a) all
excess charges are positive (b) $n_{33}=0$ (c) $n_{ij} \ge n_{i'j'}$ for $i
\ge i', j \ge j'$, we obtain from (\ref{eq:xsmatrix}):
\bea
p &=& 3(a_8+b_8)-a_3-b_3, \nonumber \\
q &=& 6(a_8+b_8). \label{eq:FNpq}
\eea
Hence the eigenvalues are simply of order
\beq
O\left( \left({\theta \over M_U} \right)^{3(a_8+b_8)+a_3+b_3} \right),
O\left( \left({\theta \over M_U} \right)^{3(a_8+b_8)-a_3-b_3} \right),
O(1).
\eeq
We will refer to this case as the {\em Froggatt-Nielsen hierarchical
structure}.

Like the fermion mass ratios, the measured quark mixing angles show a clear 
hierarchy, which is obvious in Wolfenstein's parametrization of the CKM 
matrix \cite{wolf}: 
\bea
  V_{CKM}  &  =  &  \left( \begin{array}{ccc}
	1 - \la{2} / 2  &  \lambda  &  A \la{3} (\rho + i \eta) \\
	- \lambda  &  1 - \la{2} / 2  &  A \la{2} \\
	A \la{3} (1 - \rho + i \eta)  &  - A \la{2} & 1   \end{array}  \right)
\eea
where $\lambda$ is the Cabibbo angle and $ A \simeq 0.9 \pm 0.1 $. When 
extrapolated near the Planck scale, $V_{CKM}$ keeps the same structure: the
only parameter affected by the renormalization is $A$, which is reduced by $
\simeq 30 \% $ \cite{RRR}. For our purpose, only the order of magnitude of the
mixing angles is of interest.

In order to determine the CKM matrix, we have to diagonalize both $Y^{U}$ and 
$Y^{D}$:
\bea
Diag(m_{u},m_{c},m_{t})  &  =  &  R^{U}_{L} Y^{U} R^{U \dagger}_{R}  \nonumber  \\
Diag(m_{d},m_{s},m_{b})  &  =  &  R^{D}_{L} Y^{D} R^{D \dagger}_{R}  \\
V_{CKM}  &  =  &  R^{U}_{L} R^{D \dagger}_{L}  \nonumber
\eea
This task becomes simpler if we assume that, in both charge sectors, the 
rotation matrices $R_{L}$ and $R_{R}$ can be decomposed into three small
rotations: 
\beq
R_{L} = \left(  \begin{array}{ccc}
	   1    &  -s_{12}  &  0  \\
	s_{12}  &     1     &  0  \\
	   0    &     0     &  1  \end{array}  \right)
	\left(  \begin{array}{ccc}
	   1    &     0     &  -s_{13}  \\
	   0    &     1     &     0     \\
	s_{13}  &     0     &     1     \end{array}  \right)
	\left(  \begin{array}{ccc}
	 1  &     0    &     0     \\
	 0  &     1    &  -s_{23}  \\
	 0  &  s_{23}  &     1     \end{array}  \right)
\eeq
and similarly for $R_{R}$, with rotation angles $s'_{12}$, $s'_{13}$,
and $s'_{23}$. In this parametrization, the CKM matrix reads, at leading 
order \cite{HR}:
\bea
  V_{CKM}  &  \simeq  &  \left( \begin{array}{ccc}
	1 & - s_{12} - s^{U}_{13} s_{23} & - s_{13} + s^{U}_{12} s_{23} \\
	s_{12} + s^{D}_{13} s_{23} & 1  & - s_{23} - s^{U}_{12} s_{13} \\
	s_{13} - s^{D}_{12} s_{23} & s_{23} + s^{D}_{12} s_{13} & 1          
\end{array}  \right)
\eea
where $s_{ij} = s^U_{ij} - s^{D}_{ij}$. With the additional assumption that, 
in each Yukawa matrix, the coefficient in the (3,3) entry dominates over all
other coefficients, one can express the rotation angles in terms of the Yukawa
matrix coefficients. Unfortunately, these expressions are rather complicated
\cite{HR,NLS,DPS}, unless the Yukawa matrices possess the  Froggatt-Nielsen
hierarchical structure. In this case,
\bea
  R_U^L  &  \simeq  &  \left( \begin{array}{ccc}
	O(1) & O(\epsilon_U^{2a_3}) & O(\epsilon_U^{3a_8 + a_3}) \\
	O(\epsilon_U^{2 a_3}) & O(1)  & O(\epsilon_U^{3a_8 - a_3}) \\
	O(\epsilon_U^{3a_8 + a_3}) & O(\epsilon_U^{3a_8 - a_3}) & O(1)         
\end{array}  \right) 
\eea
with $\epsilon_U = <\theta> / M_U$; and similarly for $R_L^D$ with
$\epsilon_U$ replaced by $\epsilon_D$, and $V_{CKM}$ with $\epsilon_U$ replaced 
by max($\epsilon_U$, $\epsilon_D$).

In the general case, it is more convenient for practical use to solve the
equations derived from the requirement that the matrix $R_{L} Y
R^{\dagger}_{R}$ be diagonal. The rotation angles in the (1,3) and (2,3)
sectors satisfy the following set of approximate equations: \beq \left\{ 
\begin{array}{lll}
  Y_{11} s'_{13} + Y_{12} s'_{23} - Y_{33} s_{13}   &  \simeq  &  - Y_{13}  \\
  Y_{21} s'_{13} + Y_{22} s'_{23} - Y_{33} s_{23}   &  \simeq  &  - Y_{23}  \\
  Y_{11} s_{13}  + Y_{21} s_{23}  - Y_{33} s'_{13}  &  \simeq  &  - Y_{31}  \\
  Y_{12} s_{13}  + Y_{22} s_{23}  - Y_{33} s'_{23}  &  \simeq  &  - Y_{32}
  \end{array}  \right.
\label{eq:3mixing}
\eeq
Due to the hierarchical structure of the Yukawa matrices, it is easy to solve 
these equations for a given $Y$ at leading order. The rotation angles in the
(1,2) sector have more complicated expressions, involving the rotation angles
of the two other sectors. However, when $s_{13} \leq {\cal O} (Y_{13})$ and
$s_{23} \leq {\cal O} (Y_{23})$ (this is the case for most phenomenologically
interesting Yukawa matrices), the expressions of $s_{12}$ and $s'_{12}$ reduce
to the simple form: 
\bea 
s_{12}  &  \sim  &  \frac{ Y_{11} Y_{21} + Y_{12}
Y_{22} }{ Y_{22}^{ 2} - Y_{11}^{ 2} + Y_{21}^{ 2} - Y_{12}^{ 2} }
\label{eq:(1,2)mixing}  \\ s'_{12}  &  \sim  &  \frac{ Y_{11} Y_{12} + Y_{21}
Y_{22} }{ Y_{22}^{ 2} - Y_{11}^{ 2} + Y_{12}^{ 2} - Y_{21}^{ 2} } 
\eea

Since our motivation for introducing an additional $U(1)$ symmetry with a 
chiral singlet scalar is to explain the observed hierarchies of fermion masses
and mixings, we must check that this class of models actually generates
phenomenologically viable Yukawa matrices. We will restrict ourselves here to
the quark sector, which is much more constrained than the lepton sector. We
assume that the scale $M$ is the same in both charge sectors
($M_{U}=M_{D}=M$). In order to reproduce the experimental value for the
Cabibbo angle, we also assume $<\theta>/M \simeq \lambda$. Using the result on
the hierarchy of mass eigenvalues (\ref{eq:MassHierarchy}) and the equations
(\ref{eq:3mixing}) and (\ref{eq:(1,2)mixing}) for the mixing angles, one can
search systematically for all quark Yukawa matrices $ ({\hat Y}^{U}, {\hat
Y}^{D}) $ reproducing the measured quark masses and mixing angles. They turn
out to be very few. In fact, the number of phenomenologically viable Yukawa
matrices is considerably reduced by the requirement that they originate from a
broken abelian symmetry with a chiral singlet. Indeed, the excess charges
$n_{ij}$ then satisfy the relations 
\bea 
n_{ij} + n_{kl}  &  =  &  n_{il} + n_{kj} 
\eea 
which are valid for both the charge -1/3 and +2/3 sectors, and
\bea 
n^{U}_{13} - n^{U}_{33} = n^{D}_{13} - n^{D}_{33}  &  &  n^{U}_{23} -
n^{U}_{33} = n^{D}_{23} - n^{D}_{33} 
\eea 
which relate the excess charges of
the two charge sectors. In addition, the number of negative $n_{ij}$ is
restricted by the condition $ \det {\hat Y} \neq 0 $.

In practice, we only found two sets of quark Yukawa matrices 
$ ({\hat Y}^{U}, {\hat Y}^{D}) $ reproducing the measured quark masses and
mixing angles. In the first one, ${\hat Y}^{U}$ and ${\hat Y}^{D}$ have no
supersymmetric zeroes (all excess charges are positive) and are of the form
proposed by Froggatt and Nielsen($a_3 = c_3=1/2,
b_3=3/2;a_8=5/6,b_8=7/6,c_8=1/6$):  
\beq
 \begin{array}{cc}
  n^{U} = \left( \begin{array}{rrr}
	8 & 5 & 3 \\
	7 & 4 & 2 \\
	5 & 2 & 0   \end{array}  \right)
& n^{D} = \left( \begin{array}{rrr}
	4 & 3 & 3 \\
	3 & 2 & 2 \\
	1 & 0 & 0   \end{array}  \right)  \\
 \\
  {\hat Y}^{U} \sim \left( \begin{array}{lll}
	\la{8} & \la{5} & \la{3} \\
	\la{7} & \la{4} & \la{2} \\
	\la{5} & \la{2} & 1   \end{array}  \right)
& {\hat Y}^{D} \sim \left( \begin{array}{lll}
	\la{4} & \la{3} & \la{3} \\
	\la{3} & \la{2} & \la{2} \\
	\lambda & 1 & 1   \end{array}  \right)
 \end{array}
\label{eq:solution1}
\eeq
In the second one, both ${\hat Y}^{U}$ and ${\hat Y}^{D}$ have two supersymmetric zeroes, which are filled in the way described in Subsection 2.1:
\beq
 \begin{array}{cc}
  n^{U} = \left( \begin{array}{rrr}
	8 & -1 & -3 \\
	13 & 4 & 2 \\
	11 & 2 & 0   \end{array}  \right)
& n^{D} = \left( \begin{array}{rrr}
	4 & -3 & -3 \\
	9 & 2 & 2 \\
	7 & 0 & 0   \end{array}  \right)  \\
 \\
  {\hat Y}^{U} \sim \left( \begin{array}{lll}
	\la{8} & \la{5} & \la{3} \\
	\la{13} & \la{4} & \la{2} \\
	\la{11} & \la{2} & 1   \end{array}  \right)
& {\hat Y}^{D} \sim \left( \begin{array}{lll}
	\la{4} & \la{3} & \la{3} \\
	\la{9} & \la{2} & \la{2} \\
	\la{7} & 1 & 1   \end{array}  \right)
 \end{array}
\label{eq:solution2}
\eeq
Both sets of quark Yukawa matrices (\ref{eq:solution1}) and 
(\ref{eq:solution2}), together with any phenomenologically acceptable lepton
Yukawa matrix, can be generated from an anomalous $U(1)_{X}$ with its
anomalies compensated for \`{a} la Green-Schwarz.

As written above, both (\ref{eq:solution1}) and (\ref{eq:solution2}) verify 
$ n^{U}_{33} = n^{D}_{33} = 0 $, which implies that the Yukawa couplings of
the top and the bottom quarks are of the same order at high energy: $ {\hat
Y}^{U}_{33} \sim {\hat Y}^{D}_{33} \sim 1 $. Now if we translate the down
quark excess charges by a positive integer $x$: \beq n'^{D}_{ij} = n^{D}_{ij}
+ x \eeq the down quark Yukawa matrix is simply modified by a factor $\lambda
^{x}$, keeping the same hierarchical structure: \beq
{\hat Y}'^{D} = \lambda ^{x} {\hat Y}^{D}
\eeq
However, the presence of supersymmetric zeroes in (\ref{eq:solution2})
spoils this relation for $x > 2$, so we can safely translate the 
$n^{D}_{ij}$ only by $x = 1$ or $2$. Since ${\hat Y}'^{D}$ and ${\hat Y}^{D}$
have the same eigenvalues and rotation angles, $ ({\hat Y}^{U}, {\hat Y}'^{D})
$ is still a phenomenologically viable set of quark Yukawa matrices, with $
{\hat Y}^{U}_{33} \sim 1 $ and $ {\hat Y}'^{D}_{33} \sim \lambda ^{x} $ at
high energy. As suggested by Jain and Shrock \cite{JS}, this can explain the
low-energy hierarchy between the top and bottom quark masses in a natural way,
without requiring a large $ \tan \beta $. On the contrary, the high-energy
relation $ {\hat Y}^{U}_{33} \sim {\hat Y}^{D}_{33} \sim 1 $ is compatible
with the low-energy top-bottom hierarchy only for large values of $\tan \beta$
($ \tan \beta \sim m_{t}/m_{b} $) \cite{COPW}.

\subsection{Mass scales}

The fact that the horizontal symmetry that we consider is anomalous 
has important consequences on the scale at which we might expect its breaking.

Indeed, as a result of suming over the masless states, there is a tadpole
``anomalous'' contribution to the D-term of the $U(1)_X$ anomalous symmetry. The
complete D-term reads \cite{DSW}
\beq
D_X = {g^3 M_{Pl}^2 \over 192 \pi^2} C_g + g \sum_i \phi_i \Phi^{\dagger}_i
\Phi_i  \eeq
where $g$ is the string coupling constant and $\phi_i$ is the X-charge of the
scalar field $\Phi_i$ (the tadpole term could alternatively be written
in terms of $M_{string} = g M_{Pl}$).

This provides a natural scale for the breaking of the anomalous $U(1)_X$
through a non-zero vacuum expectation value of our $\theta$ field of X-charge
$-1$ given directly in terms of the anomaly coefficient:
\beq
{<\theta^{\dagger} \theta > \over M_{Pl}^2} = {g^2 \over 192 \pi^2} C_g.
\eeq
Thus, if $C_g$ is not too large, the anomalous $U(1)$ symmetry is broken one
or two orders of magnitude below the string scale. This provides us with an
expansion parameter 
\begin{equation}
\epsilon = {|<\theta>| \over M_{Pl}}
\end{equation}
which is naturally small and not too small -- both
properties are welcome if one wants to relate this parameter with the Cabibbo
angle.


\section{The neutrino sector.} \label{sec:nom3} 

In this section, we consider generalisations of the Minimal Supersymmetric
Standard Model spectrum which include right-handed neutrinos, thus allowing
for non-zero neutrino masses and mixings. We study how the horizontal
abelian symmetry discussed above constrains the neutrino spectrum
\cite{Leontaris,GN,pierre}. For simplicity, 
we will assume only one right-handed neutrino per  family. 

Suppose that
we have three such fields, $\overline N_i$, each carrying
X-charge.  The superpotential now contains the new interaction  terms  
\beq
L_i{\overline N}_jH_u\left({\theta\over M_\nu}\right)^{p_{ij}}
+M_0{\overline N}_i{\overline N}_j\left({\theta\over M_0}\right)^{q_{ij}}
\ ,\label{eq:kll}
\eeq
multiplied by couplings of order one. The first term is
a Dirac mass term whereas the second one is a Majorana mass term and
involves the scale  $M_0$ which is some mass of the order of the GUT scale or
the string scale. In a standard $E_6$ description, the fields $\overline N_i$
may be found among the $SO(10)$ singlets or among the $SU(5)$ singlets in the
${\bf 16}$ of $SO(10)$, in which case they are part of a doublet under a
right-handed $SU(2)_R$.

We will assume here that the excess charges $p_{ij}$ and $q_{ij}$ are all 
positive and that $q_{33}$ (resp. $p_{33}$) is the smallest of the $q_{ij}$ 
(resp. $p_{ij}$) charges: $p_{ij} \ge p_{33} \ge 0$, $q_{ij} \ge q_{33} \ge 0$.
In other words, the 3-3 entry of the heavy and light neutrino mass matrices 
are dominant. We denote the X-charges of the right-handed neutrinos
by $f_0,f_3,f_8$.

For three families, the $6\times 6$ Majorana mass matrix is of the
form
\beq
\left(  \begin{array}{cc}
0 & {\cal M}\\
{\cal M}^T & {\cal M}_0 \end{array} \right)
\eeq
In the above ${\cal M}$ is the $\Delta I_w=1/2$ mass matrix with entries
not larger 
than the electroweak breaking scale, and ${\cal M}_0$ is 
the unrestricted $\Delta I_w=0$ 
mass matrix. Assuming that the order of magnitude of the $\Delta 
I_w=0$ masses is much larger than the electroweak scale, we 
obtain the generalized ``see-saw'' mechanism. 

The calculation of the light neutrino masses and mixing angles 
proceeds in two steps.
Let  $U^{}_0$ be the unitary matrix which diagonalizes the heavy
neutrino mass matrix ${\cal M}_0$, that is 
\beq
{\cal M}_0=U^{}_0D^{}_0U_0^T\ ,\label{eq:hhh}
\eeq
where $D_0$ is diagonal. The orders of magnitude of this  
matrix  are, using the invariance of the Yukawa couplings (\ref{eq:kll})
under $U(1)_X$,  
\beq
{\cal M}_0=M_0 \; {\cal O} \left( \begin{array}{ccc}
\epsilon_0^{2(f_0+f_3+f_8)}&
\epsilon_0^{2(f_0+f_8)}&
\epsilon_0^{2f_0+f_3-f_8}\\
\epsilon_0^{2(f_0+f_8)}&\epsilon_0^{2(f_0-f_3+f_8)}&\epsilon_0^{2f_0-f_3-f_8}\\
\epsilon_0^{2f_0+f_3-f_8}&\epsilon_0^{2f_0-f_3-f_8}
&\epsilon_0^{2(f_0-2f_8)} \end{array} \right) .
\eeq
where $\epsilon_0 = <\theta>/M_0$.
Its diagonalization yields the three eigenvalues
\beq
M_1=M_0{\cal O}(\epsilon_0^{2(f_0+f_3+f_8)}),\; \; M_2=M_0{\cal
O}(\epsilon_0^{2(f_0-f_3+f_8)}), \; \; M_3=
M_0{\cal O}(\epsilon_0^{2(f_0-2f_8)})\ .
\eeq
Under our assumptions the charges satisfy the 
inequalities
\beq
f_0 \ge 2f_8\ ,\qquad 3f_8 \ge |f_3| \ ,\label{eq:ineq}
\eeq
which allows to use immediately the results of section 2.4. The
diagonalizing matrix is 
\beq 
U_0=\; {\cal O} \left( \begin{array}{ccc}
1&\epsilon_0^{2|f_3|}&\epsilon_0^{3f_8+f_3}\\
\epsilon_0^{2|f_3|}&1&\epsilon_0^{3f_8-f_3}\\
\epsilon_0^{3f_8+f_3}&\epsilon_0^{3f_8-f_3}&1 \end{array} \right) \ .
\eeq
and the inverse mass matrix reads
\beq
{\cal M}_0^{-1} = {U_0}^* D_0^{-1} ({U_0}^*)^{T}
= {1 \over M_1} \; {\cal O} \left( \begin{array}{ccc}
1 & \epsilon_0^{2 f_3} & \epsilon_0^{3f_8+f_3}\\
\epsilon_0^{2f_3} & \epsilon_0^{4f_3} & \epsilon_0^{3f_8 + 3 f_3} \\
\epsilon_0^{3f_8 + f_3} & \epsilon_0^{3 f_8 + 3 f_3} & 
\epsilon_0^{2(3f_8 + f_3)} \end{array} \right).
\eeq
which is thus obtained from ${\cal M}_0$ simply by replacing $m_0$ and
$\epsilon_0$ by their respective inverses.
Then in the ``see-saw'' limit, the $3\times 3$ mass
matrix for the light neutrinos reads
\beq
{\hat Y}_{\nu} = -{\cal M}{\cal M}_0^{-1}{\cal M}^{T}\ =
-({\cal M}{U_0}^*) D_0^{-1} ({\cal M}{U_0}^*)^{T}.\label{eq:ggg}
\eeq 
The electroweak breaking mass term yields the matrix 
\beq
{\cal M}=m \; \epsilon_{\nu}^{p_{33}} \; {\cal O} \left( \begin{array}{ccc}
\epsilon_\nu^{3(d_8+f_8)+d_3+f_3}
&\epsilon_\nu^{3(d_8+f_8)+d_3-f_3}
&\epsilon_\nu^{3d_8+d_3}\\
\epsilon_\nu^{3(d_8+f_8)-d_3+f_3}
&\epsilon_\nu^{3(d_8+f_8)-d_3-f_3}
&\epsilon_\nu^{3d_8-d_3}\\
\epsilon_\nu^{3f_8+f_3}
&\epsilon_\nu^{3f_8-f_3}
&1 \end{array} \right)
\ ,\label{eq:sss}
\eeq
where $\epsilon_\nu = <\theta >/M_{\nu}$, and $m$ is
a mass of  electroweak breaking size.
We write $\epsilon_0=\epsilon_\nu^z$, with $z>0$. We find that
\beq
{\hat Y}_{\nu}={{\hat m}^2\over M_3}\; {\cal O}
\left( \begin{array}{ccc}
\epsilon_\nu^{6d_8+2d_3}
&\epsilon_\nu^{6d_8}
&\epsilon_\nu^{3d_8+d_3}\\
\epsilon_\nu^{6d_8}&
\epsilon_\nu^{6d_8-2d_3}
&\epsilon_\nu^{3d_8-d_3}\\
\epsilon_\nu^{3d_8+d_3}
&\epsilon_\nu^{3d_8-d_3}
&1  \end{array} \right)
\ ,\label{eq:ttu}
\eeq
where
\bea
\hat m &=& m \; \epsilon_{\nu}^{p_{33}} \; \; \; {\rm if}
\; \; z \le 1, \nonumber \\
\hat m &=& m \; \epsilon_{\nu}^{p_{33}} 
\; {\cal O}(\epsilon_{\nu}^{(1-z)(3f_8 + |f_3|)}) \; \; \; {\rm if}
\; \; z \ge 1, 
\eea
is the matrix whose eigenvalues yield the light neutrino masses and 
their mixing angles. It is diagonalized by the unitary matrix $U_{\nu}$:  
\beq
\hat Y_{\nu} =  U_{\nu} D_{\nu} U_{\nu}^T,
\eeq
in much the same way as the heavy  neutrino mass matrix ${\cal M}_0$. Assuming
again $3d_8 > |d_3|$, one finds
\beq
U_\nu= O \left( \begin{array}{ccc}
1&
\epsilon_\nu^{2|d_3|}&
\epsilon_\nu^{3d_8+d_3}\\
\epsilon_\nu^{2|d_3|}&1&\epsilon_\nu^{3d_8-d_3}\\
\epsilon_\nu^{3d_8+d_3}&\epsilon_\nu^{3d_8-d_3}
&1  \end{array}\right) \ .\label{eq:diaglight}
\eeq
The light neutrino masses are then 
\bea
m_{\nu_1}&=&{{\hat m}^2\over M_3}{\cal
O}(\epsilon_\nu^{2(3d_8+d_3)})\  ,\nonumber \\
m_{\nu_2}&=&{{\hat m}^2\over M_3}{\cal O}(\epsilon_\nu^{2(3d_8-d_3)})\ ,
\label{eq:light} \\
m_{\nu_3}&=&{{\hat m}^2\over M_3}\ . \nonumber
\eea
In order to obtain the  mixing matrix which appears in the
charged lepton current, we must fold this matrix with that which diagonalizes
the  charged lepton masses. If we let $\epsilon^{}_\nu=\epsilon_e^w$, with
$w>1$, the result is
\beq 
V=  O \left( \begin{array}{ccc}
1&
\epsilon_e^{2|d_3|}&
\epsilon_e^{3d_8+d_3}\\
\epsilon_e^{2|d_3|}&1&\epsilon_e^{3d_8-d_3}\\
\epsilon_e^{3d_8+d_3}&\epsilon_e^{3d_8-d_3}
&1 \end{array} \right) \ .
\eeq
When $0<w<1$, the matrix has the same form with $\epsilon_e$ replaced
by $\epsilon_\nu$. 
It is similar to the CKM matrix.
We note that its elements satisfy
\beq
V_{e\nu_\mu} V_{\mu\nu_\tau}\sim V_{e\nu_\tau}\ .\label{eq:leptmix}
\eeq

Unlike quark masses and mixing, we have little solid experimental 
information on the values of these parameters. The most compelling 
evidence for neutrino masses and mixings comes from the MSW 
interpretation of the deficit observed in various solar neutrino 
fluxes. In this picture, the electron neutrino mixes with another 
neutrino (assumed here to be the muon neutrino) 
with a mixing angle $\theta_{12}$ such that
\beq
|m^2_{\nu_1}-m^2_{\nu_2}| \sim  7\times 10^{-6}~{\rm eV^2}\ ;\; \; \; \; 
\sin^22\theta_{12}\sim  5\times 10^{-3}\ .\label{eq:solar}
\eeq

The other piece of evidence comes from the deficit of muon neutrinos 
in the collision of cosmic rays with the atmosphere.
If taken at face value, these suggest that the muon neutrinos 
oscillate into another species of neutrinos, say $\tau$ neutrinos, 
with a mixing angle $\theta_{23}$, and masses such that
\beq
|m^2_{\nu_2}-m^2_{\nu_3}| \sim 2\times 10^{-2}~{\rm eV^2}\ ;\; \; \; \; 
\sin^22\theta_{23}\ge .5  \ .\label{eq:atmos}
\eeq
Fitting the parameters coming from the solar neutrino data is rather
easy, suggesting that
\beq
V_{e\nu_\mu}\sim\epsilon_e^{2d_3}\sim \lambda^2\ ,
\eeq
together with $m_{\nu_2}\approx 10^{-3}$ eV. The atmospheric neutrino data 
would imply 
\beq
V_{\mu\nu_\tau}\sim \epsilon_e^{3d_8-d_3}={\cal O}(1)\ .
\eeq
The relation
\beq
{m_{\nu_2}\over m_{\nu_3}}\approx (V_{\mu\nu_\tau})^{2w}\ ,
\eeq
would then imply that $w>1$. For example the  
value $\theta_{23}\sim 
{\pi\over 8}$ yields $m_{\nu_2}/ m_{\nu_3}\sim .02$ , for $w=2$. Thus 
we could marginally reproduce the ``data". The 
heaviest neutrino weighs one tenth of an eV, not enough to be of use 
for structure formation.

Generically, though, it is difficult to understand 
mixing angles of order one, as suggested by the atmospheric neutrino 
data. The existence of only small mixing angles in 
the quark sectors suggests either 
that the interpretation of the atmospheric neutrino data is premature, 
or that there is fine tuning in the neutrino matrices.


\section{R-parity breaking interactions.}
\label{sec:nom4}

The gauge and Yukawa couplings are not the only interactions allowed by the 
gauge symmetries and supersymmetry. The following terms, which violate either 
B or L, can also be present in the superpotential:
\beq
\Lambda_{ijk} L_{i} L_{j} \bar e_{k} + \Lambda'_{ijk} L_{i} Q_{j} \bar d_{k} 
+ \Lambda''_{ijk} \bar d_{i} \bar d_{j} \bar u_{k}
\label{eq:RpBreaking}
\eeq
The two last ones are the most dangerous because they give rise to 
proton 
decay, if simultaneously present. In the MSSM, R-parity is assumed in order 
to forbid them. We consider here the most general case where R-parity may be 
broken, and we therefore take into account all these terms (the $L_{i} H_{u}$ 
term, which one usually eliminate by a redefinition of $H_{d}$, will be 
discussed later on). The couplings $\Lambda_{ijk}$, $\Lambda'_{ijk}$ and 
$\Lambda''_{ijk}$ must then be very small, otherwise they would induce proton 
decay and lepton number violation at an unacceptable level. The upper bounds, 
due to the experimental limits on respectively proton decay, lepton number 
violation and neutron-antineutron oscillations, are \cite{HK}:
\bea 
\sqrt{\Lambda' \Lambda''}  &  \leq  &  {\cal O} \left[ \left( {M_{Susy} \over 
1 TeV} \right) 10^{-13} \right]  \label{eq:pDecay}  \\
\Lambda  &  \leq  &  {\cal O} \left[ \left( {M_{Susy} \over 1 TeV} \right) 
10^{-3} \right]  \label{eq:LViolation}  \\
\Lambda''  &  \leq  &  {\cal O} \left[ \left( {M_{Susy} \over 1 TeV} 
\right)^{5/2} 10^{-5} \right] 
\label{eq:nOscillation}
\eea
Note that the most stringent constraint comes from proton decay 
(\ref{eq:pDecay}). It is satisfied if one of the two terms $L Q \bar d$ or 
$\bar d \bar d \bar u$ is highly suppressed, or if both are. Another 
possibility is that either $L Q \bar d$ or $\bar d \bar d \bar u$ do not 
appear in the superpotential. In the following, we shall look at both 
possibilities.

The horizontal symmetry $U(1)_{X}$ discussed in the previous sections 
naturally generates small couplings \cite{BN}. Let us consider, for example, 
the $L_{i} Q_{j} \bar d_{k}$ term. It carries the excess charge 
$x_{ijk} = l_{i} + q_{j} + d_{k}$. If $x_{ijk} > 0$, $L_{i} Q_{j} \bar d_{k}$ 
will be generated from the non-renormalizable interaction:
\beq
a_{ijk} L_{i} Q_{j} \bar d_{k} \left( {\theta \over M} \right) ^{x_{ijk}}
\eeq
where $a_{ijk}$ is a factor of order one. The effective $\Lambda_{ijk}$ 
coupling will then be of order $(<\theta> / M)^{x_{ijk}}$. If $x_{ijk} < 0$, 
the $L_{i} Q_{j} \bar d_{k}$ term will not appear in the superpotential. But, 
in the same way as the Yukawa couplings whose excess charges are negative 
(see subsection 2.1), it can be induced by non-renormalizable contributions 
to the kinetic terms. The effective $\Lambda_{ijk}$ coupling is then:
\beq
\Lambda_{ijk} = \sum_{l,m,n} H(x_{lmn}) \Lambda_{ijk;lmn}
\eeq
where $\Lambda_{ijk;lmn}$ is the contribution of the nonzero $L_{l} Q_{m} 
\bar d_{n}$ term to the $L_{i} Q_{j} \bar d_{k}$ term. It is given by:
\beq
\Lambda_{ijk;lmn} \sim \left( {<\theta> \over M} \right) ^{|l_{l}-l_{i}| 
+ |q_{m}-q_{j}| + |d_{n}-d_{k}| + x_{lmn}}
\label{eq:contribution}
\eeq
One deduces from (\ref{eq:contribution}) that $\Lambda_{ijk}$ is at most of 
the order of magnitude that would be obtained with a vectorlike pair of 
$\theta$ fields:
\beq
\Lambda_{ijk} \leq {\cal O} \left( \left( {<\theta> \over M} 
\right) ^{|x_{ijk}|} \right)
\label{eq:ZeroFilling}
\eeq
If $x_{ijk} \geq 0$, this bound is saturated because $\Lambda_{ijk;ijk} = 1$. 
Therefore, the diagonalization of the kinetic terms does not affect the order 
of the couplings which are initially nonzero.

The only difference with the Yukawa couplings is that the number of negative 
excess charges is not limited by the condition of requiring a nonzero 
determinant: a single positive $x_{ijk}$ is then sufficient to generate all 
other R-parity violating terms of the same type. However, this mechanism 
tends to produce small couplings. For example, in the particular case where 
there is a single positive excess charge $x_{lmn}$, one can easily show that 
the $\Lambda_{ijk}$ induced by the diagonalization of the kinetic terms are 
of the order of:
\beq
\Lambda_{ijk} \sim \left( {<\theta> \over M} \right) ^{|x_{ijk}| + 2 x_{lmn}}
\eeq
while, of course, $\Lambda_{lmn} \sim (<\theta> / M)^{x_{lmn}}$. If $x_{lmn}$ is large enough, this leads to very small couplings. This property holds when there are several positive $x_{ijk}$, provided that all of them are large compared to unity.

We conclude that, in order to obtain small R-parity violating couplings, we 
must choose the X-charges of the MSSM fields so that all positive $x_{ijk}$ 
are large. The number of negative $x_{ijk}$ does not matter; the important 
point is that the smallest positive excess charge be large. Thus all 
effective $\Lambda_{ijk}$ will be small. We require that all of them be very 
small, because the physical couplings, which enter the proton decay rate, 
involve mass eigenstates and therefore mix the $\Lambda_{ijk}$.
 This mixing tends to attenuate the hierarchy between R-parity violating 
couplings of the same type (say $L Q \bar d$), in disagreement with what 
is usually assumed in phenomenological analysis.

In practice, it is not so easy to obtain large positive excess charges for 
the $\Lambda_{ijk}$. Indeed, the family-dependent part of the X-charge is 
very constrained by the quark phenomenology, and its family-independent part 
is fixed by the Green-Schwarz compensation of its anomalies. These constraints
disfavor large values of the $x_{ijk}$. The only freedom we have, provided 
that the neutrinos are massless, is to choose the lepton charges. 
Unfortunately, they must have very large
values, which seems to be rather unnatural.

This is shown by the following example, where $Y_{U}$ and $Y_{D}$ have the 
form proposed by Froggatt and Nielsen \cite{FN}. The charge assignment is the 
following:

\newpage
{ \bf Table 1:} X-charges of the MSSM fields according to the family
index $i = 1, 2 ,3$  (first example).
\vskip 1cm
\begin{tabular}{||c|c|c|c|c|c|c||}
\hline
& & & & & & \\
$i$ & $q_i$ &$u_i$ &$d_i$ & $l_i$ & $e_i$ & $h_u = h_d$ \\
& & & & & & \\
\hline
& & & & & & \\
1 & 2/3 & 22/3 & 16/3 & -12 & 18 &  \\
& & & & & & \\
\cline{1-6}
& & & & & & \\
2 & -1/3 &13/3 &13/3 &-13 &17 &0\\
& & & & & & \\
\cline{1-6}
& & & & & & \\
3 &-7/3 &7/3 &13/3 &55 &-53 &\\
& & & & & & \\
\hline
\end{tabular}
\vskip 1.5cm

The corresponding Yukawa matrices are:
\bea
  Y_{U} \sim \left( \begin{array}{lll}
	\la{8} & \la{5} & \la{3} \\
	\la{7} & \la{4} & \la{2} \\
	\la{5} & \la{2} & 1   \end{array}  \right)
& Y_{D} \sim \la{2} \left( \begin{array}{lll}
	\la{4} & \la{3} & \la{3} \\
	\la{3} & \la{2} & \la{2} \\
	\lambda & 1 & 1   \end{array}  \right)  \nonumber
\eea
\beq
  Y_{E} \sim \la{2} \left( \begin{array}{lll}
	\la{4} & \la{3} & \la{67} \\
	\la{3} & \la{2} & \la{68} \\
	\la{71} & \la{70} & 1   \end{array}  \right)
\label{eq:example1}
\eeq
where $\lambda = (<\theta> / M)$ is assumed to be the Cabibbo angle. The 
constraints (\ref{eq:pDecay}) to (\ref{eq:nOscillation}) are widely satisfied 
by a strong suppression of L violation: 
\bea
\Lambda  &  \leq  &  {\cal O} \left( 10^{-38} \right) \nonumber \\
\Lambda'  &  \leq  &  {\cal O} \left( 10^{-38} \right) \\
\Lambda''  &  \leq  &  {\cal O} \left( 10^{-7} \right) \nonumber
\eea
but, as stressed above, the lepton charges are large, which gives rise to 
very small coefficients in the lepton Yukawa matrix. Ben-Hamo and Nir 
\cite{BN} did not encounter this problem because they did not consider the 
anomalies of $U(1)_{X}$.

As mentioned above, another possibility for avoiding proton decay is that 
one of the two dangerous terms $L Q \bar d$ and $\bar d \bar d \bar u$ be 
absent from the superpotential. This happens when all excess charges for this 
term are negative, because all corresponding couplings are then zero. For 
example, one can find a large class of $U(1)_{X}$ models, in which there is 
no $\bar d \bar d \bar u$ term. These models are interesting, because the 
experimental constraints then reduce to (\ref{eq:LViolation}), which is very 
easy to satisfy. Unfortunately, they also have very large values for the 
lepton charges.

Our second example belongs to this class of models. The charge assignment is 
the following:

\vskip .8cm
{ \bf Table 2:} X-charges of the MSSM fields according to the family
index $i = 1, 2 ,3$  (second example).
\vskip 1cm
\begin{tabular}{||c|c|c|c|c|c|c||}
\hline
& & & & & & \\
i & $q_i$ &$u_i$ &$d_i$ & $l_i$ & $e_i$ & $h_u=h_d$ \\
& & & & & & \\
\hline
& & & & & & \\
1 & 23/3 & 1/3 & -5/3 &  23 & -17 & \\
& & & & & & \\
\cline{1-6}
& & & & & & \\
2 & 20/3 & -8/3 &-8/3 & 22 & -18 &0\\
& & & & & & \\
\cline{1-6}
& & & & & & \\
3 &14/3 & -14/3 &-8/3 &-78 & 80 &\\
& & & & & & \\
\hline
\end{tabular}
\vskip 1.5cm
The corresponding Yukawa matrices are:
\bea
  Y_{U} \sim \left( \begin{array}{lll}
	\la{8} & \la{5} & \la{3} \\
	\la{7} & \la{4} & \la{2} \\
	\la{5} & \la{2} & 1   \end{array}  \right)
& Y_{D} \sim \la{2} \left( \begin{array}{lll}
	\la{4} & \la{3} & \la{3} \\
	\la{3} & \la{2} & \la{2} \\
	\lambda & 1 & 1   \end{array}  \right)  \nonumber
\eea
\beq
  Y_{E} \sim \la{2} \left( \begin{array}{lll}
	\la{4} & \la{3} & \la{101} \\
	\la{3} & \la{2} & \la{100} \\
	\la{97} & \la{98} & 1   \end{array}  \right)
\label{eq:example2}
\eeq
As stressed above, there is no B violation from renormalizable operators, 
and the remaining constraint (\ref{eq:LViolation}) is widely satisfied:
\bea
\Lambda  &  \leq  &  {\cal O} \left( 10^{-16} \right) \nonumber \\
\Lambda'  &  \leq  &  {\cal O} \left( 10^{-16} \right) \\
\Lambda''  &  =  &  0 \nonumber
\eea

We must also consider the possibility that the $x_{ijk}$ be fractionnary, 
which is generally the case. The effective couplings are then zero, unless 
they are due to non-perturbative effects. Indeed, if one of the $x_{ijk}$ is 
fractionnary, all of them are fractionnary. All $\Lambda_{ijk}$ are then 
initially zero, and remain zero after diagonalization of the kinetic terms. 
This follows from the fact that the excess charges of the Yukawa couplings 
are integers. Consider now the three terms in (\ref{eq:RpBreaking}). One can 
easily show that the excess charges of the first two terms are simultaneously 
fractionnary or integers, while the excess charges of the third term can be 
fractionnary or integers independently from the first two terms. We can 
therefore choose the lepton charges so that only the L violating terms 
(resp. only the B violating term) are present in the superpotential, which 
makes proton decay impossible in the absence of higher dimension operators.

So far, we did not consider the higher dimension R-parity violating operators.
Two of them give a significant contribution to proton decay \cite{HK,BN}:
\beq
{\kappa_{ijkl} \over M} Q_{i} Q_{j} Q_{k} L_{l}  + {\kappa'_{ijkl} \over M} 
\bar u_{i} \bar u_{j} \bar d_{k} \bar e_{l}
\eeq
The upper bounds on the $\kappa$ couplings are:
\bea
\kappa  &  \leq  &  {\cal O} \left[ \left( {M_{Susy} \over 1 TeV} \right) 
\left( {M \over M_{P}} \right) 10^{-7} \right]  \\
\kappa' (K^{U}_{RR})_{1i}  &  \leq  &  {\cal O} \left[ \left( {M_{Susy} 
\over 1 TeV} \right) \left( {M \over M_{P}} \right) 10^{-8} \right]
\eea
where $K^{U}_{RR}$ is the quark-squark mixing matrix for the right-handed up 
quarks. When there is no mixing (no FCNC), $K^{U}_{RR} \equiv 0$ and there is 
no constraint over $\kappa'$. These constraints are easily satisfied as soon 
as the lepton charges are large. This is the case as well in the first 
example (\ref{eq:example1}): 
\bea 
\kappa  &  \leq  &  {\cal O} \left( 10^{-32} \right)  \nonumber  \\
\kappa' &  \leq  &  {\cal O} \left( 10^{-17} \right)
\eea
as in the second one (\ref{eq:example2}):
\bea
\kappa  &  \leq  &  {\cal O} \left( 10^{-24} \right)  \nonumber  \\
\kappa' &  \leq  &  {\cal O} \left( 10^{-45} \right)
\eea

In the previous discussion, we did not mention the $L H_{u}$ term, which 
should be present in the superpotential, in addition to the three terms of 
(\ref{eq:RpBreaking}). One usually eliminate it by a redefinition of $H_{d}$. 
Starting with the following quadratic part of the superpotential:
\beq
\mu H_{u} H_{d} + \alpha_{i} L_{i} H_{u}
\eeq
and redefining $H'_{d} = H_{d} + \sum_{i} (\alpha_{i} / \mu) L_{i}$, one ends 
up with a single quadratic term, $\mu H_{u} H'_{d}$. It is important to note 
that, in our model, this can be done only after the breaking of $U(1)_{X}$, 
because the $H_{d}$ and $L_{i}$ superfields carry different X-charges. Now 
the redefinition of $H_{d}$ also modifies the Yukawa terms of the down quarks:
\bea
\lambda^{D}_{jk} H_{d} Q_{j} \bar d_{k}  &  \rightarrow  &  
\lambda^{D}_{jk} H'_{d} Q_{j} \bar d_{k} - \left( {\alpha_{i} \over \mu} 
\right) \lambda^{D}_{jk} L_{i} Q_{j} \bar d_{k} 
\eea
which gives a new contribution to the $L_{i} Q_{j} \bar d_{k}$ term (in a 
similar way, $L_{i} L_{j} \bar e_{k}$ receives a contribution from 
$H_{d} L_{j} \bar e_{k}$). The effective $\Lambda_{ijk}$ is then modified as 
follows:
\bea
\Lambda_{ijk}  &  \rightarrow  &  \Lambda_{ijk} 
+ \left( {\alpha_{i} \over \mu} \right) \lambda^{D}_{jk}
\eea
Thus the $L_i H_{u}$ term, if present, contributes to the L violating 
couplings, and we must take it into account in our analysis. Note that, 
since the $\alpha_{i}$ are generated in the same way as the $\Lambda_{ijk}$, 
they are zero as soon as the excess charges $(l_{i} + h_u)$ are fractionnary 
or all negative. In this case, the $L H_{u}$ term does not appear in the 
superpotential. Otherwise, it may give 
their dominant contribution to the $\Lambda_{ijk}$. In particular, when the 
excess charges of the $L L \bar e$ and $L Q \bar d$ terms are fractionnary, 
only $L H_{u}$ contributes to the L violating couplings.

We can distinguish between two cases:
\begin{enumerate}
\item  if $L L \bar e$ and $L Q \bar d$ have fractionnary excess charges, the 
L violating couplings $\Lambda$ and $\Lambda'$ are generated from the 
$L H_{u}$ term. However, when $h_u \in Z$, $L H_{u}$ is absent, and there is 
no L violation from renormalizable operators.
\item  if $L L \bar e$ and $L Q \bar d$ have integer excess charges, the 
$L H_{u}$ term is present only if $h_u \in Z$. When $h_u=0$ however, its 
contribution does not modify the order of magnitude of the L violating 
couplings (this is the case in both examples given). 
\end{enumerate}

\section{Conclusions} \label{sec:nom5}

Trying to explain fermion mass hierarchies and mixings by an {\em ad hoc}
local abelian gauge symmetry might seem, at first glance, an honest but
somewhat groundless attempt. Surprisingly, this leads to a very special
type of abelian symmetry, namely the  anomalous $U(1)$  whose anomalies may be
cancelled by the Green-Schwarz mechanism. This leaves some hope that, in the
context of string models, one may be able to make definite statements about
mass hierarchies. Indeed, because of the uniqueness of the dilaton field, such a
$U(1)$ symmetry is  unique and plays a central r\^ole. One may therefore
relate the charges of the matter fields under this $U(1)$ to central properties
of the model. Such a $U(1)$ has already been advocated \cite{faraggi} to
explain why a nonvanishing top Yukawa coupling may appear at string tree level.
Its properties may also allow to relate the horizontal symmetry approach to the
modular symmetries of the underlying string theory \cite{BD}.

Surprisingly little information from the anomaly structure of this  symmetry
is used to derive the Weinberg angle -- or, in a correlated way, the order of
magnitude of the $\mu$ term in a certain class of models --. One may expect
that the rest of the information, in particular the mixed gravitational
anomaly which plays a r\^ole in fixing the scale at which this symmetry
breaks, can be used to constrain further the models \cite{SB}.

We have also studied two types of extended supersymmetric standard
models -- massive neutrinos and R-parity breaking interactions --, where this
approach  proves to be (mildly) constraining. It is for instance interesting to
see that, when trying to implement in this framework a generalized seesaw
mechanism for neutrinos, one ends up with a light neutrino mass spectrum which
cannot satisfy at the same time the cosmological and atmospheric neutrino
constraints.

\vskip 1cm 
{\bf Acknowledgements}
\vskip .5cm
P.B. wishes to thank for hospitality the Institute for Theoretical Physics in
Santa Barbara where part of this work was done. P.R.   thanks Mr
N. Irges for numerous discussions during the course of this work.

\end{document}